\documentclass[12pt]{article}
\usepackage[dvips]{graphicx}
\usepackage[latin1]{inputenc}
\newcommand{\ii} {\'{\i}}
\author{%
Huitzilin Y\'epez${}^{1}$\footnote{E-mail: hyepez@fis.cinvestav.mx}, 
Juan M. Romero${}^{2}$\footnote{E-mail: sanpedro@nuclecu.unam.mx}
and Adolfo Zamora${}^{2}$\footnote{E-mail: zamora@nuclecu.unam.mx}\\
\\
${}^{1}${\it Departamento de F\ii sica, Cinvestav-IPN,}\\
 {\it Apartado Postal 14-740-543, M\'exico DF, M\'exico.}\\
\\
${}^{2}${\it Instituto de Ciencias Nucleares, UNAM,}\\
 {\it Apartado Postal 70-543, M\'exico DF, M\'exico.}\\}

\usepackage[latin1]{inputenc}
\date{}

\begin{document}

\title{Corrections to the Planck's radiation law from 
loop quantum gravity\\}

\maketitle

\begin{abstract}
We study the dispersion relation obtained from the semiclassical
loop quantum gravity. This dispersion relation is considered for
a photon system at finite temperatures and the changes to the 
Planck's radiation law, the Wien and Boltzmann laws are discussed. 
Corrections to the equation of state of the black body radiation
are also obtained.
\end{abstract}

Pacs: 04.60.Pp, 05.70.-a, 98.70.Ve 

\newpage

\section{Introduction}

Recently it has been studied with great interest the possibility
that particles undergo changes in their dispersion relation as
a consequence of gravitational or high energy effects. Hawking's 
spectrum for black holes, for instance, displays changes in the
range of high frequencies \cite{jaco:gnus}. An analogous effect
appears in the inflationary cosmology for the inflaton spectrum
\cite{martin:gnus}. Another example is found in the field theory 
of non-commutative spaces, where the dispersion relation for 
photons also changes \cite{jackiw:gnus}.
In the semiclassical loop quantum gravity (SLQG) it is possible 
to find an expression for a field over the weak quantum gravitational
background. For this field corrections to its dispersion relation 
are also obtained \cite{urrutia1:gnus,urrutia2:gnus}. For a review 
see Ref. \cite{hugo:gnus}. The modified dispersion relation could 
give rise to several consequences as the breaking of the Lorentz 
symmetry, for example.
As an alternative approach, modified dispersion relations have 
been used to provide an explanation of anomalies in ultrahigh 
energetic cosmic rays \cite{amelino:gnus}.\\

In general one would expect that the consequences of changes to a
dispersion relation were observed only in highly energetic particles. 
However, this is not always the case. To see this let us consider 
the modified dispersion relation for photons,
\begin{eqnarray}
p^{2}=E^{2}\left[1+\xi E/E_{p}+\mathcal{O}((E/E_{p})^{2})\right],
\label{eq:beli}
\end{eqnarray}
where $\xi\approx 1$ and $E_{p}\approx 10^{18}$ $GeV$ is the Planck
energy. In this case, a photon that travels a distance $L$ will
have a time delay
\begin{eqnarray}
\Delta t\approx\xi \frac{L}{c}(E/E_{p}).
\label{eq:beli2}
\end{eqnarray}
It can be seen from Eq. (\ref{eq:beli2}) that if the distance $L$
is large enough, this time delay will be observed even if the energy 
of the particle is not large. In other words, this is a small 
microscopic effect that could be observed at a macroscopic scale 
provided the macroscopic variables are large enough. For the 
case of the SLQG, the corrections to the dispersion relation for 
photons are analogous to those described by Eq. (\ref{eq:beli}).\\

A well known result from condensed matter is that small changes in 
the microscopic properties of a system can lead to important 
modifications in its macroscopic behavior. For quantum liquids, for 
example, it can be seen that a change in the dispersion relation of 
the excitations can result in a completely different behavior. Liquid 
helium, ${}^{4}$He, provides a good example of a boson fluid with a 
small residual interaction. At low temperatures this interaction is 
the one responsible for the superfluid state \cite{landau:gnus}. The 
main excitations below $1K$ are just phonons with a dispersion relation 
$\omega\propto k$, but at higher temperatures roton excitations with
$\omega\propto k^2$ must be included. Accounting for these, corrections 
to the thermodynamic properties can be observed 
\cite{landau:gnus,feynman:gnus}.\\

In this letter we study the changes to the macroscopic properties at 
finite temperatures of a photon system provided the dispersion relation 
proposed by the SLQG is assumed. To first order in the Planck length, 
the dispersion relation is linear in the polarization and so the 
thermodynamic properties remain unchanged. However, to second order, 
the Planck's radiation law loses its universal character as it will be 
shown shortly. Consequences of losing this universal behavior, as it 
is a correction to the Wien's displacement law, are also discussed. 
We finally find that the Boltzmann law for the radiated energy density 
and the radiation pressure change.\\

This work is organized as follows. In section $2$ the dispersion 
relation obtained from the SLQG is presented. It is also shown here 
that the black body radiation law loses its universal behavior and 
that this leads to modifications to the Wien's law. In section $3$ 
the changes to the thermodynamic properties are obtained; and finally 
in section $4$ our results are summarized.

\section{The dispersion relation from the LQG}

The dispersion relation obtained from the SLQG for photons is given 
by \cite{urrutia1:gnus,hugo:gnus}:
\begin{eqnarray}
\omega=ck+\lambda a_{1} k^{2}-a_{2}k^{3},
\label{eq:al10}
\end{eqnarray}
where $a_{1}$ is proportional to the Planck length $l_{p}$, $a_{2}$ 
is proportional to $l_{p}^{2}$ and $\lambda$ is the polarization of 
the photons that can take values $\lambda=\pm 1$. The coefficients 
$a_{1}$ and $a_{2}$ depend on the semiclassical state chosen in the 
LQG formalism.\\

It is worth mentioning that Eq. (\ref{eq:al10}) has been obtained 
under the hypothesis that $kl_{p}\ll 1$; which forbids the solution 
$k_{c}\ne 0$ for $\omega=0$. Notice that such a $k_{c}$ is of the 
order of
\begin{eqnarray}
k_{c} \propto \frac{1}{l_{p}},
\end{eqnarray}
which is very large and does not satisfy $kl_{p}\ll 1$. Therefore, 
moments $k\ge k_{c}$ can be neglected. It can also be observed that 
there is a value $k_{max}$ for which $\omega$ from Eq. (\ref{eq:al10})
has its maximum. This $k_{max}$ is of the same order of magnitude as
$k_{c}$.\\

\begin{figure}[tbh]
\begin{center}
\includegraphics[width=12.0 cm,height=9.0cm]{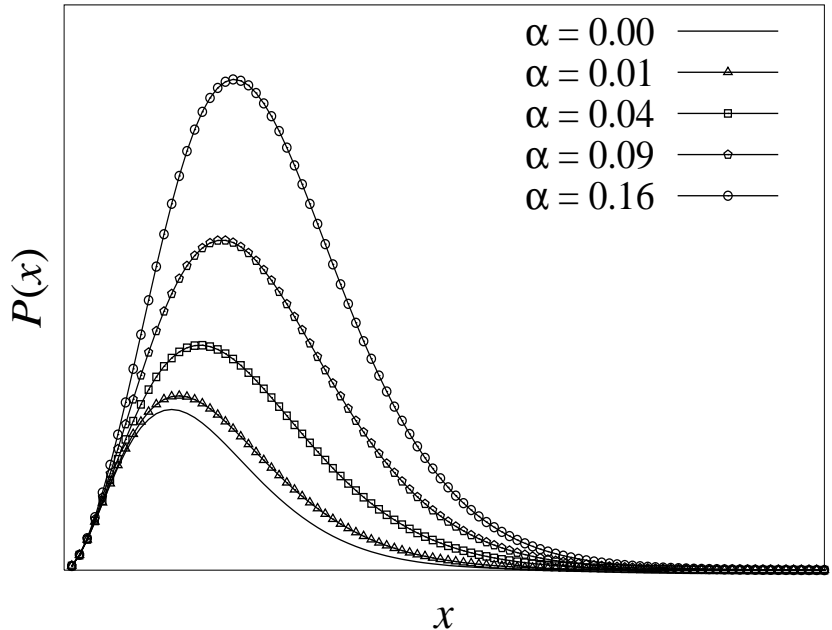}
\end{center}
\caption{{\protect\small Modified Planck distribution for different 
values of temperature [$\alpha = 0$ corresponds to $P_0(x)$.
$\alpha = 0.01,0.04,0.09,0.16$ are for $P(x)$ at $T = 1,2,3,4$
in units of $100 C_2 k_B^2/\hbar^2$.]}}
\label{teq}
\end{figure}

Let us now proceed to calculate the modified Planck's radiation law. 
The number of particles inside a volume element in phase space is
\begin{eqnarray}
dN_{\omega}=\rho(\omega)\frac{4\pi V}{(2\pi)^{3}}k^{2}dk,
\end{eqnarray}
where $V$ is the volume and $\rho(\omega)$ is given by
\begin{eqnarray}
\rho(\omega)=\frac{1}{e^{\beta\hbar\omega}-1},
\end{eqnarray}
with $\beta=(k_{B}T)^{-1}$, $k_{B}$ being the Boltzmann constant and 
$T$ the temperature. The energy associated with the above volume 
element is
\begin{eqnarray}
dU_{\omega}=\hbar\omega\rho(\omega)\frac{V}{2\pi^{2}}k^{2}dk.
\label{eq:al200}
\end{eqnarray}
Using the dispersion relation in Eq. (\ref{eq:al10}) to first order 
in $l_{p}$ we obtain for the momentum $k$
\begin{eqnarray}
k=\frac{\omega}{c}\left[1-\frac{\lambda a_{1}}{c^{2}} \omega\right].
\end{eqnarray}
However, if considered to second order, we get
\begin{eqnarray}
k=\frac{\omega}{c}\left[1-\frac{\lambda a_{1}}{c^{2}} \omega+
\left(\frac{2a_{1}^{2}}{c^{4}}+\frac{a_{2}}{c^{3}}\right)
\omega^{2}\right],
\end{eqnarray}
so that, to this order
\begin{eqnarray}
k^{2}dk=\frac{\omega^{2}}{c^{3}}[1-\lambda
C_{1}\omega+C_{2}\omega^{2}]\, d\omega,
\end{eqnarray}
with
\begin{eqnarray}
C_{1}=\frac{4a_{1}}{c^{2}}\quad \mathrm{and} \quad 
C_{2}=\frac{15a_{1}^{2}}{c^{4}}+\frac{5a_{2}}{c^{3}}.
\label{eq:al21}
\end{eqnarray}
Therefore, the energy in Eq. (\ref{eq:al200}) becomes
\begin{eqnarray}
dU_{\omega}=\frac{\hbar V}{2\pi^{2}} \frac{\omega^{3}}{c^{3}}%
\rho(\omega)[1-\lambda C_{1}\omega+C_{2}\omega^{2}]\, d\omega.
\label{eq:al300}
\end{eqnarray}
Notice, in this equation, that after summing over all possible 
$\lambda$ values, the corrections to first order in $l_{p}$ get 
canceled. Therefore the first corrections to the thermodynamic 
quantities appear to second order. That is
\begin{eqnarray}
dU_{\omega }&=&\frac{V\hbar }{\pi ^{2}c^{3}}\rho (\omega )\omega
^{3}[1+C_{2}\omega ^{2}]\, d\omega  \nonumber \\
&=&\frac{V\hbar }{\pi ^{2}c^{3}(\hbar \beta )^{4}}\frac{x^{3}}{e^{x}-1}
\left[1+\frac{C_{2}}{(\hbar \beta )^{2}}x^{2}\right]\, dx,
\label{eq:al1}
\end{eqnarray}
with $x=\hbar\omega\beta$. To order zero Eq. (\ref{eq:al1}) yields
the usual Planck distribution,
\begin{eqnarray}
P_{0}(x)=\frac{x^{3}}{e^{x}-1},
\label{eq:amo}
\end{eqnarray}
which is valid for the radiation of a black body at any given 
temperature and therefore it is universal. However, the first
corrections to Eq. (\ref{eq:al1}) yield the modified distribution
\begin{eqnarray}
P(x)=\frac{x^{3}}{e^{x}-1}(1+\alpha x^{2}),\quad 
\alpha=\frac{C_{2}k_{B}^{2}}{\hbar^{2}}T^{2},
\label{eq:al3}
\end{eqnarray}
that does depend on temperature and thus it loses the universal
property observed in $P_{0}(x)$. Figure \ref{teq} shows the behavior 
of the modified Planck distribution $P(x)$ from Eq. (\ref{eq:al3}) 
compared to the usual Planck distribution $P_{0}(x)$ for different 
temperatures.

Another way to express the universal behavior of $P_{0}(x)$ can be 
obtained by considering that it is invariant under the transformation
\begin{eqnarray}
\omega \to \gamma \omega ,\qquad T\to \gamma T.
\label{eq:chido}
\end{eqnarray}
If Eq. (\ref{eq:chido}) is assumed, to zero order in $l_{p}$, the 
energy $dU_{\omega}$ scales with $\gamma^{4}$. That is, it preserves 
form but presents changes in its global scale. Including the quadratic 
corrections, $dU_{\omega}$ changes both its form and scale.\\

A consequence of the breaking of universality is that the maximum of
$P(x)$ depends on temperature. For the usual Planck distribution,
$P_{0}(x)$ the maximum occurs when
\begin{eqnarray}
x_{0max}=\frac{\hbar\omega_{max}}{k_{B}T}\simeq 2.822, 
\label{eq:wien1}
\end{eqnarray}
which is invariant under the transformation in Eq. (\ref{eq:chido}). 
The expression for $x_{0max}$ in Eq. (\ref{eq:wien1}) is the well 
known  Wien's displacement law: the frequency for the maximum in 
the Planck distribution variates linearly with temperature. However, 
for $P(x)$ the maximum occurs when $x$ satisfies the condition
\begin{eqnarray}
(3-x)e^{x}-3+\alpha x^{2}(5e^{x} -xe^{x} -5)=0.
\label{eq:al11}
\end{eqnarray}
To first order in $\alpha$, the solution is given by
\begin{eqnarray}
x_{max}\simeq x_{0max}+(18.218)\alpha.
\label{eq:maximum}
\end{eqnarray}
Notice that $x_{max}$ does not have the symmetry of Eq. (\ref{eq:chido})
because of its temperature dependence. From Eq. (\ref{eq:maximum}) we 
get
\begin{eqnarray}
\omega_{max}=\left(\frac{2.822k_{B}}{\hbar}\right)T+
\left(\frac{18.218 C_{2}k_{B}^{3}}{\hbar^{3}}\right) T^{3},
\label{eq:wien2}
\end{eqnarray}
with $C_{2}$ defined in Eq. (\ref{eq:al21}). Therefore, the Wien's 
displacement law gets a correction proportional to $T^{3}$.\\

It is clear from Wien's law, Eq. (\ref{eq:wien1}), that if an 
astrophysical object radiates with a well known frequency, then it 
will be possible to know its temperature. However, from the modified 
Wien's law, Eq. (\ref{eq:wien2}), we notice that the expression for 
$\omega_{max}$ is cubic in $T$ and no longer linear. Nevertheless, 
these cubic corrections are usually very small so they can only be 
observed at high temperatures.\\

Now, we investigate the cosmological consequences of losing universality 
in Eq. (\ref{eq:al1}). If the Robertson-Walker metric is considered, 
the frequency and temperature for the scaling are both proportional to 
the inverse radius of the universe $R$ \cite{carroll:gnus},
\begin{eqnarray}
\omega \propto \frac{1}{R}, \qquad T \propto \frac{1}{R}.
\label{eq:chi1}
\end{eqnarray}
By assuming this scaling transformation it can be seen that the
energy $dU_{\omega}$ from Eq. (\ref{eq:al1}) to order zero preserves
form but changes scale with expansion of the universe. However, once
the corrections are included, it displays modifications in both
form and scale.\\

The dependence on $R$ of the temperature is derived from the 
Boltzmann law. We will see in the next section that this law 
is modified. This last result changes the dependence on $R$ 
of $T$, but does not solve the problem of loosing the universal 
character of $P(x)$.\\

There are theories from which one can also get corrections to the 
Planck distribution \cite{manko:gnus,minic:gnus}. Some of these do 
not actually break universality of the distribution. As an example, 
in quantum groups theory, the Planck distribution obtained has a 
complex structure but the universal character remains \cite{manko:gnus}. 
However, if a quantum mechanics consistent with the existence of a 
minimal length is considered, the Planck distribution loses 
universality \cite{minic:gnus}.

\section{Changes in the thermodynamic properties}

The energy of the photon system can be obtained by integrating 
out Eq. (\ref{eq:al1}) and evaluating the upper limit at the 
momentum $k_{max}$ from section $2$. As the momentum $k_{max}$
is very large and the energy decreases rapidly for 
$\omega >\omega_{max}$, we can approximate the value of the 
energy by the integral over all possible values of $k$.
Considering these simplifications we find
\begin{eqnarray}
U=\frac{4V}{c}T^{4}\sigma (T),
\label{eq:al2}
\end{eqnarray}
with $\sigma(T)$ given by
\begin{eqnarray}
\sigma(T)=\sigma_{0}\left[1+\left(\frac{40C_{2}k_{B}^{2}\pi^{2}}
{21\hbar^{2}}\right) T^{2}\right],\quad \sigma_{0}=\frac{\pi^{2}
k_{B}^{4}}{60\hbar^{3}c^{2}}.
\label{eq:al5}
\end{eqnarray}
Equation (\ref{eq:al2}) represents the modified Boltzmann law. 
The correction obtained appears in the Stephan-Boltzmann constant, 
Eq. (\ref{eq:al5}), which now displays a quadratic dependence on 
temperature.\\

Using Eq. (\ref{eq:al2}) we obtain for the specific heat at 
constant volume,
\begin{eqnarray}
C_{V}=\frac{16VT^{3}}{c}\sigma_{0}\left[1+
\left(\frac{60C_{2}k_{B}^{2}\pi^{2}}{21\hbar^{2}}\right)T^{2}\right].
\end{eqnarray}
To get other thermodynamic quantities it is necessary to know not
only the energy $U$ but also the partition function. In the grand 
canonical ensemble, the grand potential for bosons is given by 
\cite{reichl:gnus}
\begin{eqnarray}
\Omega=k_{B}
T\sum_{l>0}\sum_{\lambda}\ln(1-e^{-\beta(\epsilon_{l\lambda}-\mu)}),
\end{eqnarray}
where $l$ corresponds to the $l$-th momentum and $\lambda$ is the
polarization. Additionally, for the case of photons $\mu =0$. The
sum above can be replaced by an integral with aid of the rule
\begin{eqnarray}
\sum_{l} \quad \to \quad \frac{V}{(2\pi)^{3}}\int dk^{3}=
\frac{V4\pi}{(2\pi)^{3}}\int k^{2} dk.
\end{eqnarray}
Performing these substitutions we obtain
\begin{eqnarray}
&\Omega&=\frac{\kappa TV}{2\pi^{2}c^{3}} \sum_{\lambda}\int_{0}^{\infty}
\ln(1-e^{-\beta\hbar\omega_{\lambda}}) \omega^{2}_{\lambda}[1-\lambda
C_{1}\omega_{\lambda}+C_{2}\omega_{\lambda}^{2}]d\omega_{\lambda}  \nonumber
\\
& &=\frac{\kappa TV}{\pi^{2}c^{3}}\int_{0}^{\infty}\ln(1-e^{-\beta\hbar
\omega)}) \omega^{2}[1+C_{2}\omega^{2}]d\omega  \nonumber \\
& &=-\left(\frac{V\kappa^{4} \pi^{2}}{45\hbar^{3}c^{3}}\right)T^{4}\left[1+
\left(\frac{8C_{2}\kappa^{2}\pi^{2}}{7\hbar^{2}}\right)T^{2}\right],
\label{eq:a500}
\end{eqnarray}
so that the entropy is
\begin{eqnarray}
S=-\left(\frac{\partial\Omega}{\partial T}\right)_{V}=\left(\frac{%
4\kappa^{4} \pi^{2}}{45\hbar^{3}c^{3}}\right)T^{3}\left[1+\left(\frac{%
12C_{2}\kappa^{2}\pi^{2}}{7\hbar^{2}}\right)T^{2}\right].
\end{eqnarray}
With this result, considering Eq. (\ref{eq:a500}), and using the
thermodynamic relation
\begin{eqnarray}
U=\Omega+TS,
\end{eqnarray}
we recover Eq. (\ref{eq:al2}) for the energy $U$. In addition, for 
the radiation pressure we obtain
\begin{eqnarray}
P=\left(\frac{k_{B}^{4} \pi^{2}}{45\hbar^{3}c^{3}}\right)T^{4}
\left[1+\left(\frac{8C_{2}k_{B}^{2}\pi^{2}}{7\hbar^{2}}\right)
T^{2}\right].
\label{eq:feli}
\end{eqnarray}
From this it can be observed that the radiation pressure has
a small increase respect to the standard result appreciable
only at high temperatures.\\

The equation of state for the radiation pressure and energy density 
is usually expressed in the form
\begin{eqnarray}
P=\bar\omega \frac{U}{V}.
\end{eqnarray}
In general, $\bar\omega$ is constant, but in this case it has a
temperature dependence of the form
\begin{eqnarray}
\bar\omega =\frac{1}{3}\left[1-\left(\frac{16C_{2}k_{B}^{2} \pi^{2}}
{21\hbar^{2}}\right)T^{2}\right].
\end{eqnarray}
Notice that the correction to the usual equation of state is 
negative, which is analogous to the equation of state for 
a space with cosmological constant \cite{carroll:gnus}.

\section{Summary}

In this work we consider the dispersion relation for photons 
obtained from the SLQG and study the thermodynamic changes.
One of these changes is that the modified Planck distribution 
does not preserve universality. This is because the corrections 
depend on frequency but not on temperature and therefore it is 
not possible to express the modified distribution in terms of 
a dimensionless variable. Some consequences of this lack of 
universality were reviewed. One of these is the corrections to 
the Wien's displacement law. Here, there is a small shift of the 
maximum frequency of the Planck distribution which depends on 
the cubic temperature. Boltzmann's law and the radiation pressure 
also get small corrections. The equation of state for the energy 
density and pressure also gets modified. The constant relating 
pressure and energy density now has negative corrections that 
depend quadratically on temperature.

\end{document}